\documentclass{desyproc}

\begin{document}
%------------------------------------
\title{Physics with first LHCb data}

% for the author list please adhere to the format of one of the following
% three examples

% use the following for a single author
%
%\author{{\slshape Joe Smith}\\[1ex]
%DESY, Notketra{\ss}e 85, 22607 Hamburg, Germany }

% use the following for several authors
%
%\author{{\slshape Jean Meunier$^1$, Ruth Miller$^2$,
%    Gerd M\"uller$^3$\footnote{Speaker}, Joe Smith$^3$}\\[1ex]
%$^1$CERN, 1211 Gen\`eve 23, Switzerland\\
%$^2$Fermilab, P.O. Box 500, Batavia, IL 60510-0500, USA\\
%$^3$DESY, Notketra{\ss}e 85, 22607 Hamburg, Germany}

% use the following for an author speaking on behalf of a collaboration
%
\author{{\slshape Olivier Schneider}  for the LHCb Collaboration\\[1ex]
Ecole Polytechnique F\'{e}d\'{e}rale de Lausanne (EPFL), CH--1015 Lausanne, Switzerland}

% please do not modify the following 5 lines
\contribID{xy}  % will be entered by the editors
\confID{1964}
\desyproc{DESY-PROC-2010-01}
\acronym{PLHC2010}
\doi            % will be entered by the editors

\maketitle

\begin{abstract} % max. 100 words
The LHCb experiment is designed for hadronic flavour physics 
and will look for New Physics manifestations in the decay of charm and
bottom hadrons abundantly produced at the LHC. All parts of the LHCb physics 
programme can be embarked on with the expected statistics 
to be collected during the first 
2010--2011 physics run at $\sqrt{s}=7$~TeV.
We present first
preliminary results on strangeness production, and demonstrate, 
using the few nb$^{-1}$ of already collected data, 
the potential for initial measurements in heavy-flavour physics.
\end{abstract}

\section{Physics goals and strategy}

The Standard Model (SM) of particle physics cannot be the ultimate theory. 
It is incomplete and contains too many free parameters, 
such as masses and quark mixing angles. The pattern of these parameters 
should be governed by a hidden mechanism yet to be discovered, and so the 
SM is believed to be a low-energy effective theory of a 
more fundamental theory at a higher energy scale, anticipated to be in 
the TeV region and accessible at the Large Hadron Collider (LHC). 
This would imply new symmetries, particles, dynamics, 
and flavour structure. 

The most exciting task of the LHC experiments will be to 
find this New Physics, whatever it may be. 
This can be done either directly or 
indirectly. The direct approach, pursued mostly by the 
ATLAS and CMS experiments, aims at the observation of new 
particles produced in LHC's proton-proton collisions at 14~TeV. 
The indirect approach, on the other hand, consists in measuring 
quantum corrections in the decay of already known particles 
especially in flavour-changing neutral-current (FCNC) transitions, 
and looking for deviations from the SM predictions.
At LHC, this will be best done by the LHCb experiment, which has been designed 
specifically for precise measurements of CP violation and rare decays 
of hadrons containing a $b$ quark. Both approaches are complementary:
while the indirect approach is sensitive 
to higher energy scales and may therefore sense a new effect earlier, 
the direct observation of any new particle 
is necessary to establish its unambiguous discovery as well as 
for measuring its main properties.
New Physics (NP) 
at the TeV scale needs to have a non-trivial flavour structure 
in order to provide the suppression mechanism for the already observed
FCNC processes. 
Only indirect measurements can access the 
phases of the new couplings and therefore shed light 
on the NP flavour structure. 

One of the strategies for indirect searches in hadronic decays 
consists of measuring
as many observables as possible that can be related 
to the magnitudes and phases of 
the elements of the Cabibbo-Kobayashi-Maskawa (CKM) matrix describing the 
SM flavour structure in the quark sector. 
Any inconsistency between the interpretation of these measurements within
the CKM picture will be a sign of New Physics. 
The most awaited progress in this area is a precise NP-free determination 
of the CKM angle $\gamma$ from tree-level processes. 

Another strategy is to identify and measure single FCNC processes
with good NP discovery potential, {\em i.e.} where NP is likely to
emerge and for which a clear SM prediction can be made. 
Decays involving the $b\to s$ transition, which is less constrained
by the current data, are good candidates.
They are theoretically calculated using the Operator Product Expansion 
in terms of short-distance Wilson coefficients 
and long-distance operators describing effective vertexes such as
tree diagrams, 
or gluon-, photon-, electroweak-, scalar- and pseudoscalar-penguin loops.
New physics may both enhance some of 
the Wilson coefficients or introduce new operators, in particular in the 
right-handed sector which is suppressed in the SM. 

Following these strategies, LHCb
is preparing to perform rate measurements
(such as the $B^0_s\to\mu^+\mu^-$ branching fraction), 
determine CP-violating phases (most notably mixing-induced 
effects in $B^0_s\to J/\psi\phi$ and $B^0_s\to\phi\phi$
decays, interference between $b\to u$ and $b \to c$ transitions in 
tree-level $B\to D K$ decays, CP asymmetries in 
charmless two-body $B$ decays), and probe the helicity structure 
of weak interactions (photon polarization in $B^0_s\to\phi\gamma$ and other 
radiative decays, asymmetries in $B^0\to K^{*0}\mu^+\mu^-$ decays). 
Such promising measurements are central to the core physics programme of LHCb;
they have been studied in detail and are described in 
a recent roadmap document~\cite{LHCb:2009ny}.
However, the wider programme will include many more measurements, 
mostly in (but not limited to) the heavy-flavor sector. 

\begin{figure}[t]
\begin{center}
\epsfig{file=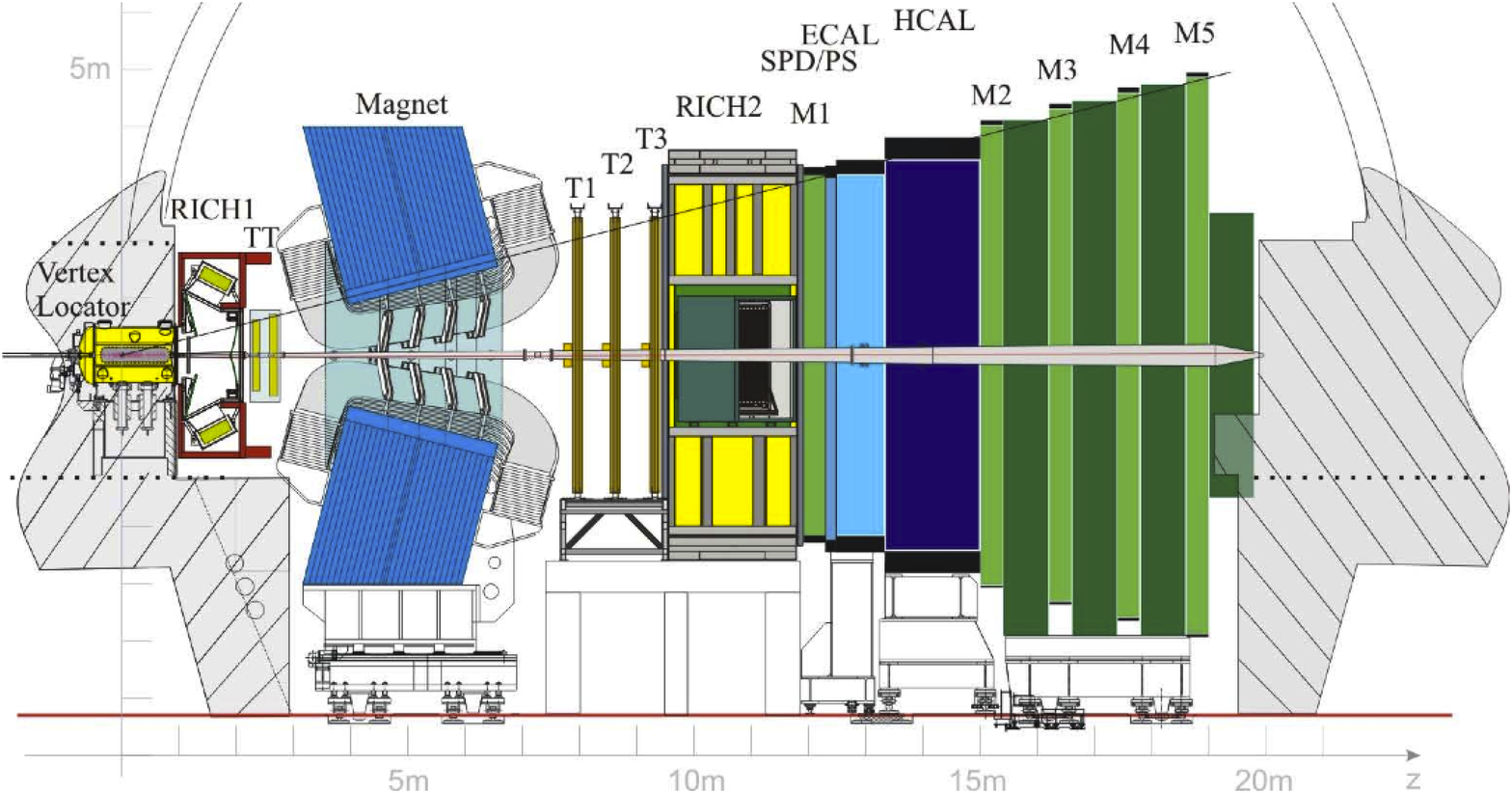,width=0.49\textwidth}
\raisebox{1ex}{\epsfig{file=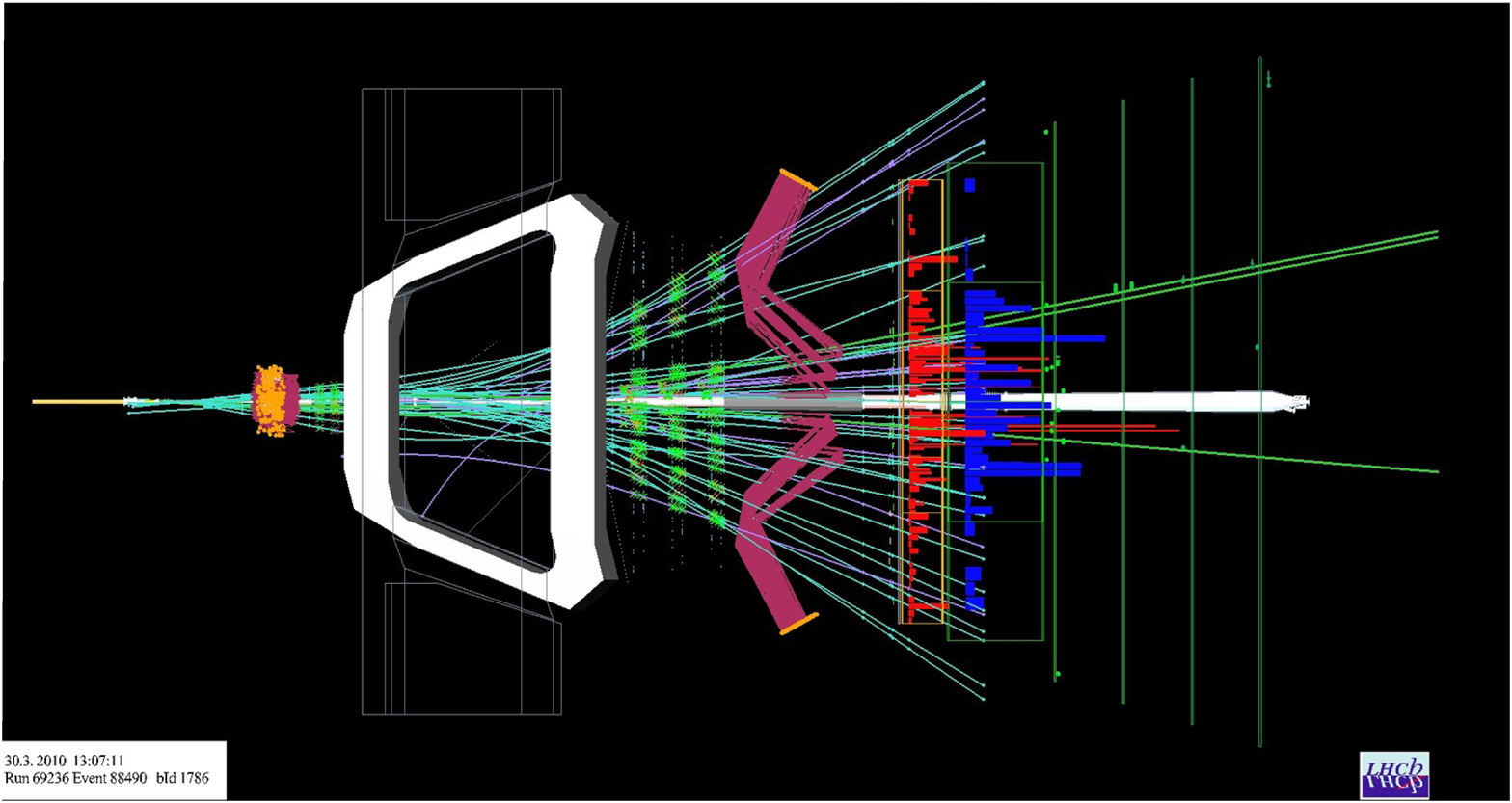,width=0.49\textwidth}}
\caption{\underline{Left:}
Side view of the LHCb spectrometer, showing the Vertex Locator
around the interaction region on the left, 
the tracking stations before (TT) and after (T1--T3) the dipole magnet,
the ring-imaging Cherenkov detectors (RICH1 and RICH2), 
the calorimeter system (SPD/PS, ECAL, HCAL), and the muon stations (M1--M5).
\underline{Right:} Event display (top view, in bending plane)
of one of the first recorded $pp$ collisions at 
$\sqrt{s}=7$~TeV on March 30, 2010. Reconstructed tracks,
originating from the $pp$ collision point on the left,
have been reconstructed from hits in the VELO and hits (green)
in the tracking stations. Cherenkov photons (mauve) are reflected
on mirrors towards photo-detectors (orange).
Energy depositions in ECAL (red) and HCAL (blue)
as well as hits in the muon chambers (green, far right) are also visible.}
\label{fig:LHCb}
\end{center}
\end{figure}

\section{LHCb and first physics run}

The LHCb detector~\cite{Alves:2008zz} is a single-arm spectrometer
(see Fig.~\ref{fig:LHCb} left) 
covering the forward region ($1.9 < \eta < 4.9$)
where the $b\bar{b}$ production is peaked.
It will rely on relatively soft $p_{\rm T}$ triggers, 
efficient for both leptonic $B$ decays ($\sim 90\%$) 
and purely hadronic $B$ decays ($\sim 40\%$).
By design the luminosity will be limited
to an average of $\rm \sim 2 \times 10^{32}~cm^{-2}s^{-1}$
in order to avoid a significant fraction of events with
more than one $pp$ inelastic interaction.
A nominal year ($\rm 10^7~s$) of running in design conditions will give 
an integrated luminosity of $\rm 2~fb^{-1}$ at 
$\sqrt{s} = \rm 14~TeV$. However, in the first LHC physics run 
started on March 30, 2010 (see Fig.~\ref{fig:LHCb} right)
the centre-of-mass energy is $\sqrt{s} = \rm 7~TeV$, reducing 
the expected $b\bar{b}$ and $c\bar{c}$ production rates by factors
$\sim 2.3$ and $\sim 1.8$, respectively, 
% This factor was obtained from the bb cross section given by PYTHIA
% with the LHCb tuning, 
% see page https://twiki.cern.ch/twiki/bin/view/LHCb/SettingsMc09:
% sigma(bb) = 1.040 +- 0.01 mb at sqrt(s) = 14 TeV
% sigma(bb) = 0.529 +- 0.01 mb at sqrt(s) =  8 TeV
% sigma(bb) = 0.380 +- 0.01 mb at sqrt(s) =  6 TeV
% A linear interpolation at gives:
% sigma(bb) = (0.529+0.380)/2 = 0.455 mb at sqrt(s) = 7 TeV
% The ratio to 14 teV is then:
% 1.040/ 0.455 = 2.29
although without dramatic impact on the physics reach.
The nominal instantaneous luminosity is expected
to be reached in 2011, 
while the current lower luminosity period in 2010
allows for lower trigger thresholds, and hence 
better efficiencies for hadronic $B$ decays ($\sim 75\%$).
This represents also a good opportunity to 
collect rapidly very large samples of charm events,
with a corresponding trigger efficiency boosted up
from $\sim 10\%$ to $\sim 40\%$. 
Approximately $\rm 14~nb^{-1}$ of data have been collected 
during April and May 2010, mostly with a fully inclusive trigger 
requesting at least one reconstructed track in the detector. 
Since the last week of May, a loose High-Level Trigger  is run 
in rejection mode to limit the output rate to a few kHz. 
The overall status of the experiment~\cite{Golutvin_talk},
the data-taking experience~\cite{Wiedner_talk}, 
and the event reconstruction performance~\cite{Maciuc_talk,Xing_talk}
obtained from the first data are described elsewhere. 

\begin{wrapfigure}{r}{0.5\textwidth}
\begin{center}
\epsfig{file=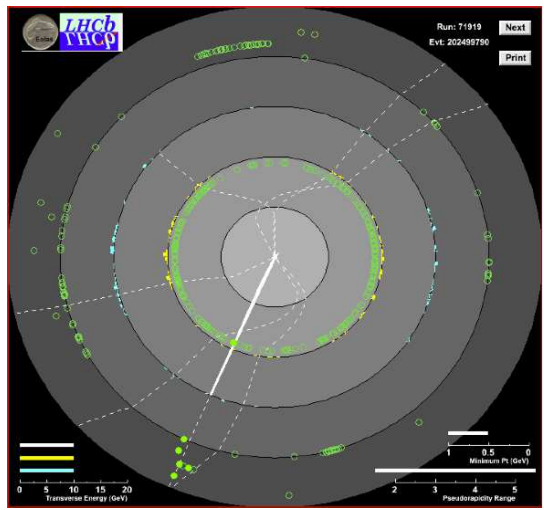,width=0.49\textwidth}
\end{center}
\caption{LHCb's first $W^+\to \mu^+\nu$ candidate,
shown in a `$z-\phi$ view' of the detector, with the Vertex Locator
at the centre and muon stations at the periphery of the display.
The white thick straight line represents a high $p_{\rm T}$ track
($p_{\rm T}=35.4$~GeV) with hits in the muon chambers,
while the curved dotted lines are accompanying soft tracks.}
\label{fig:W}
\end{wrapfigure}

The first physics measurements within reach are those of the production of 
known and most abundantly produced particles. LHCb is focusing initially on 
unstable particles which can be reconstructed through their decay into charged 
tracks, and therefore cleanly identified as narrow signals above some
combinatorial background. So far close to 30 different mass peaks 
have been seen in the LHCb data, including decays involving neutrals such as 
$\eta \to \pi^+\pi^-\pi^0$, $\eta' \to \pi^+\pi^-\gamma$, and 
$D^0 \to K^-\pi^+\pi^0$. Because of the nature of the
LHCb core measurements, which will most often rely on fully reconstructed
decays, the understanding and modelling of the structure of minimum bias
events is not of utmost importance, hence more difficult production
measurements of stable particles such as charged pions, kaons, protons
or tracks in general are not at the centre of the present effort.
Of more direct interest are the production measurements of strange
(and neutral), charm, and bottom hadrons, 
as well as of electroweak bosons (see Fig.~\ref{fig:W}). 

Production measurements at LHCb are necessarily new 
since LHC is operating at an unexplored energy.
In order to turn them into cross section measurements, an estimate of the 
luminosity is needed. 
The principle of a direct determination of the luminosity based on a
new `beam imaging' technique~\cite{FerroLuzzi:2005em}
has been demonstrated using the data collected during the LHC pilot run
in December 2009~\cite{Hopchev:2010tj}, and used for the first absolute
production cross section measurement described below. 

\section{Results on strangeness production}

\begin{figure}[t]
\begin{center}
\begin{minipage}[c]{0.52\textwidth}
\raisebox{1.6ex}{
\epsfig{file=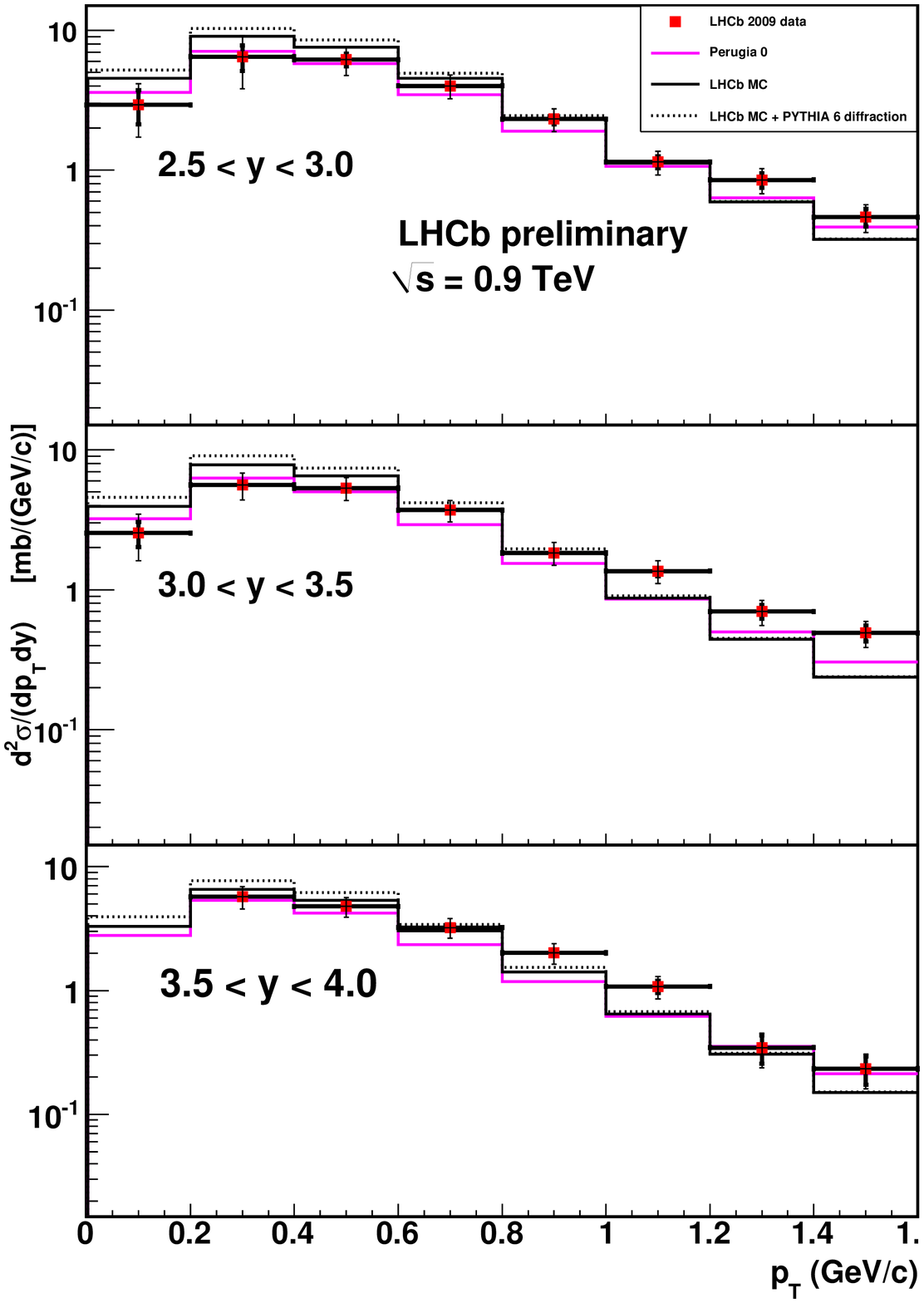,width=\textwidth}}
\end{minipage}
\begin{minipage}[c]{0.47\textwidth}
\begin{center}
\epsfig{file=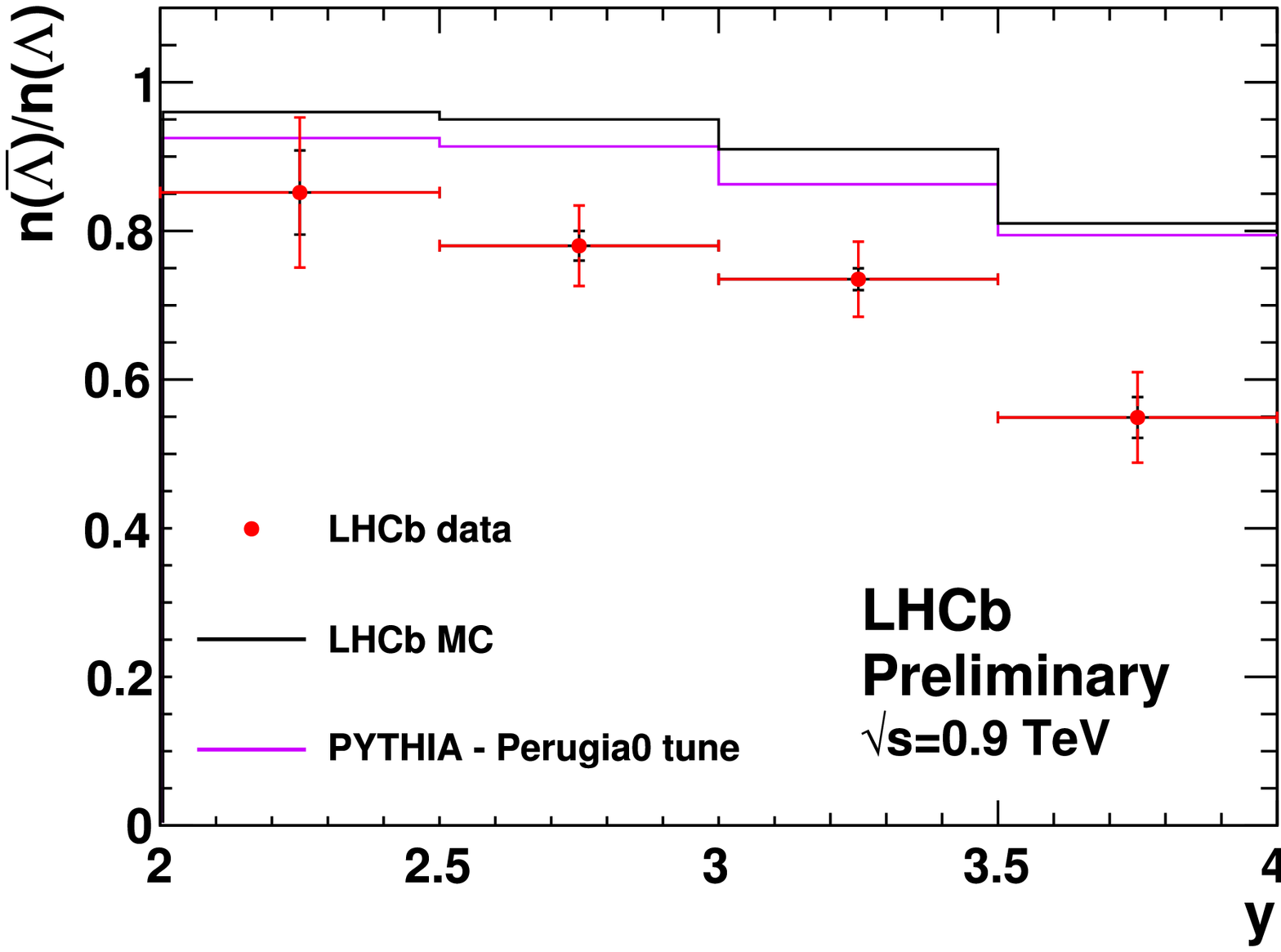,width=0.70\textwidth,angle=270}\\[0.5ex]
\epsfig{file=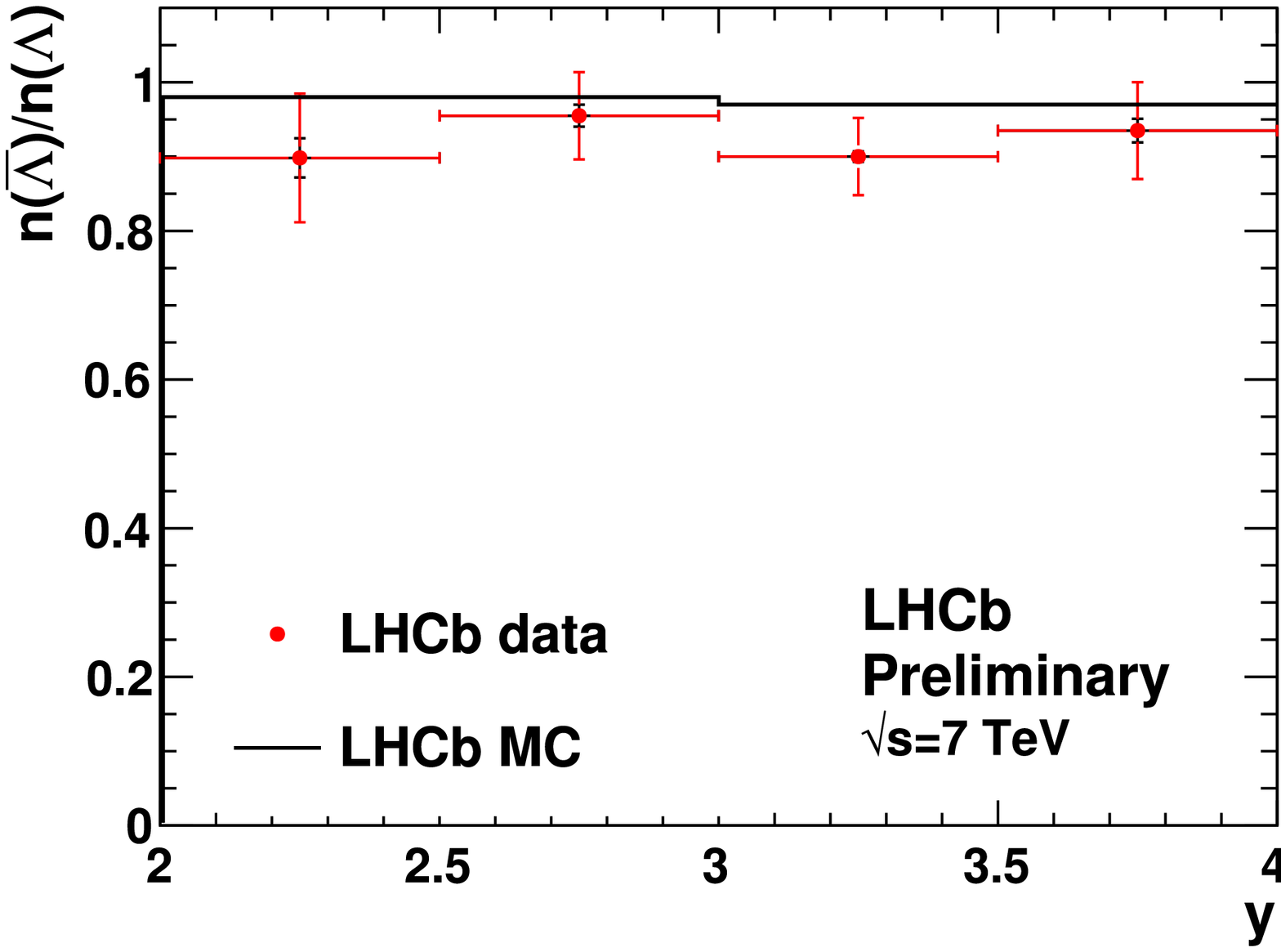,width=0.70\textwidth,angle=270}
\end{center}
\end{minipage}
\caption{\underline{Left:} 
Double-differential prompt $K^0_{\rm S}$ production cross section
in $pp$ collisions at $\sqrt{s}=0.9$~TeV, shown as a function of
$K^0_{\rm S}$ transverse momentum $p_{\rm T}$ in
three different bins in $K^0_{\rm S}$ rapidity $y$.
\underline{Right:}
$\bar{\Lambda}/\Lambda$ production ratio as a function of rapidity $y$
in $pp$ collisions
at $\sqrt{s}=0.9$~TeV (top) and 7~TeV (bottom). 
In all cases the red points represent LHCb data, with statistical
and total uncertainties shown as vertical error bars,
while the histograms are expectations from 
the PYTHIA~6.4 generator with different parameter settings,
including the LHCb Monte Carlo (black)
and the `Perugia~0' tune~\cite{Skands_Perugia} (purple).}
\label{fig:Vzero}
\end{center}
\end{figure}

Strange quarks appear in the hadronization process of soft hadronic
interactions, and their production is an excellent probe of the
fragmentation field. In particular the measurement of strangeness
production in hadronic interactions provides input for the understanding
of QCD in the non-perturbative regime and for the tuning of Monte Carlo
generators. 

The data collected during the LHC pilot run in December 2009 at
$\sqrt{s}=0.9$~TeV were used to measure the prompt $K^0_{\rm S}$
production as a function of the $K^0_{\rm S}$ transverse momentum
$p_{\rm T}$ and rapidity $y$ in the region 
$0 < p_{\rm T}<1.6~{\rm GeV}/c$ and $2.5<y<4.0$ (see Fig.~\ref{fig:Vzero} left).
At this low beam energy the beam sizes and crossing angle (induced by the
LHCb dipole magnet) do not allow the complete closure of the Vertex Locator
(VELO) around the interaction region. As a result the data were collected
with the VELO silicon detectors retracted by 15~mm from their nominal position,
reducing significantly the azimuthal coverage provided by the VELO.
However $K^0_{\rm S}\to\pi^+\pi^-$ decays could still efficiently be
reconstructed using tracks reconstructed in the tracking stations
(TT and T1--T3). On the other hand the VELO was essential to measure $pp$
and beam-gas interaction vertices, and determine the positions, 
sizes and angles of the colliding proton bunches.
Together with bunch current measurements obtained from the LHC machine
instrumentation, this allowed a direct determination of the integrated
luminosity ($6.8 \pm 1.0~\mu{\rm b}^{-1}$) of the sample used for the
$K^0_{\rm S}$ analysis.
As can be seen from Fig.~\ref{fig:Vzero} (left), the preliminary measurements 
of the absolute prompt $K^0_{\rm S}$ production cross section are in fair
agreement with the expectations from the PYTHIA generator, 
before any tuning to LHC data. These results have been finalized
and published~\cite{Kshort_paper} since the conference.

The data collected in 2010, both at $\sqrt{s}=0.9$~TeV and 7~TeV,
were also used to study $\Lambda \to p \pi^-$ production.
We show for the first time at this conference~\cite{Bonivento_talk}
preliminary measurements of the $\bar{\Lambda}/\Lambda$ production ratio
as a function of  rapidity $y$ for the two centre-of-mass energies
(Fig.~\ref{fig:Vzero} right). Contrary to the results at high energy, 
the measurements of the $\bar{\Lambda}/\Lambda$ ratio 
at $\sqrt{s}=0.9$~TeV are significantly below the expectation
and show a strong dependence in rapidity. Such studies are useful
to investigate and understand the baryon-number transport
from the beams in the more central region of the detector.

\section{Charm: first look and prospects}

\begin{figure}[t]
\begin{center}
\epsfig{file=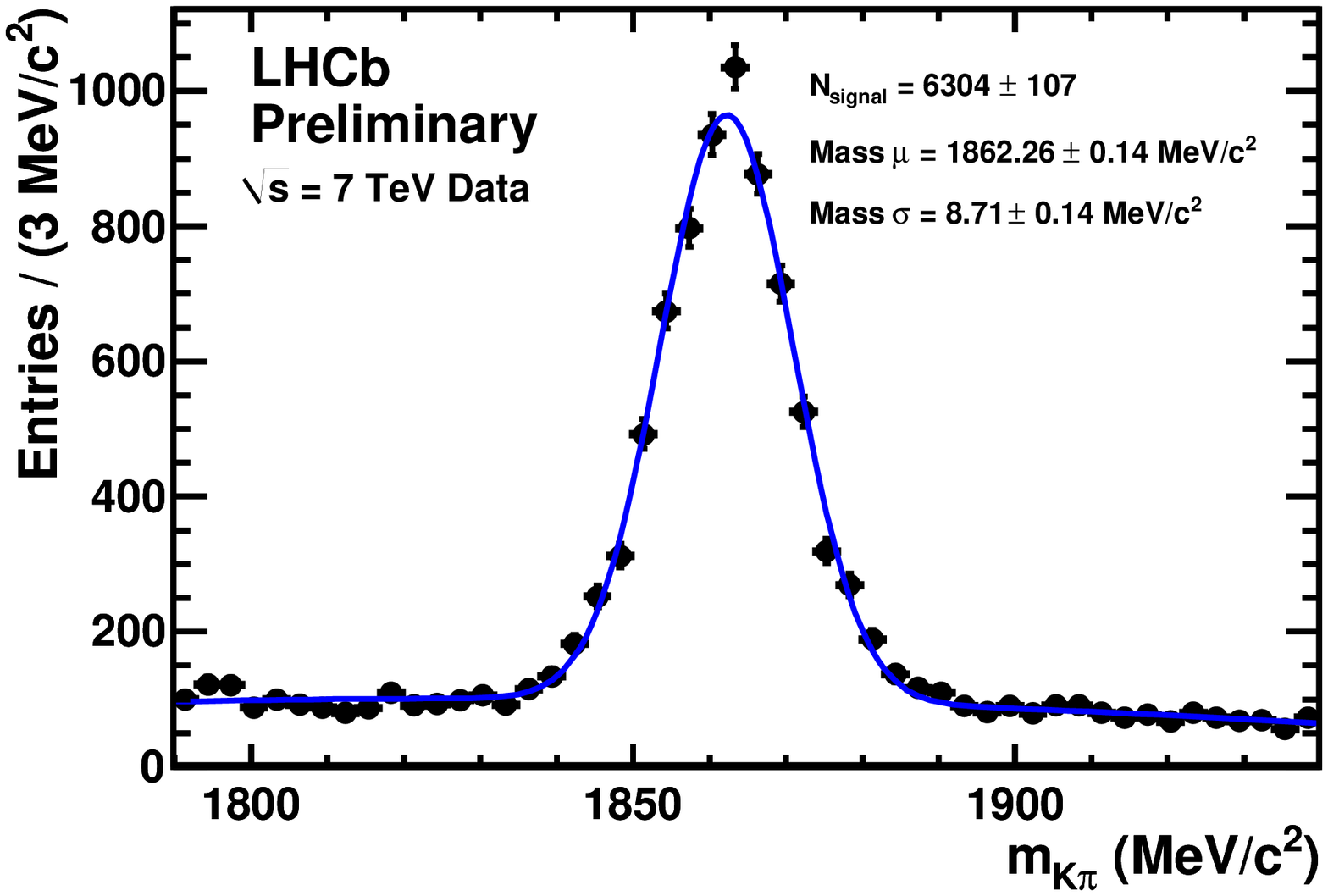,width=0.49\textwidth}
\epsfig{file=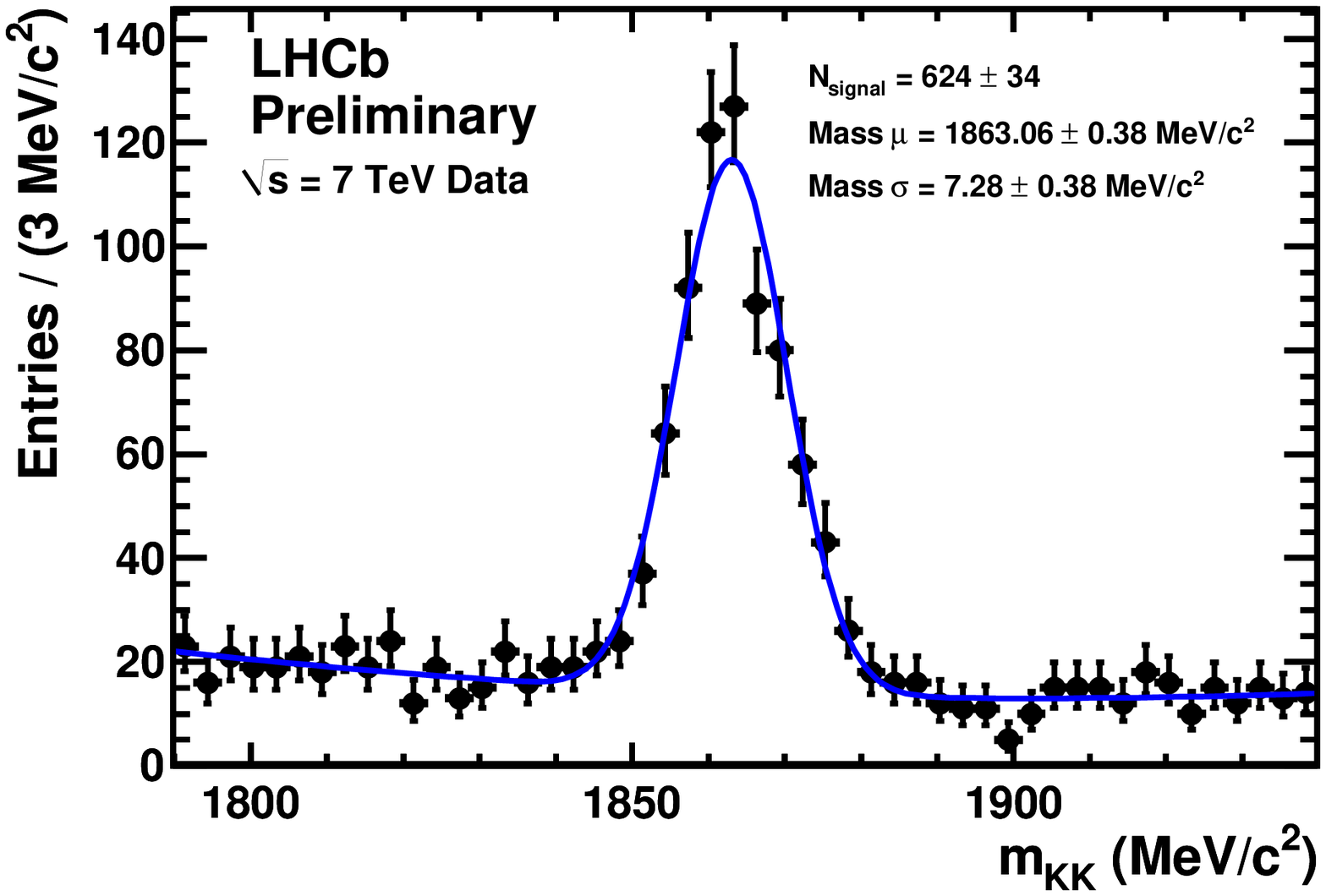,width=0.49\textwidth}
\epsfig{file=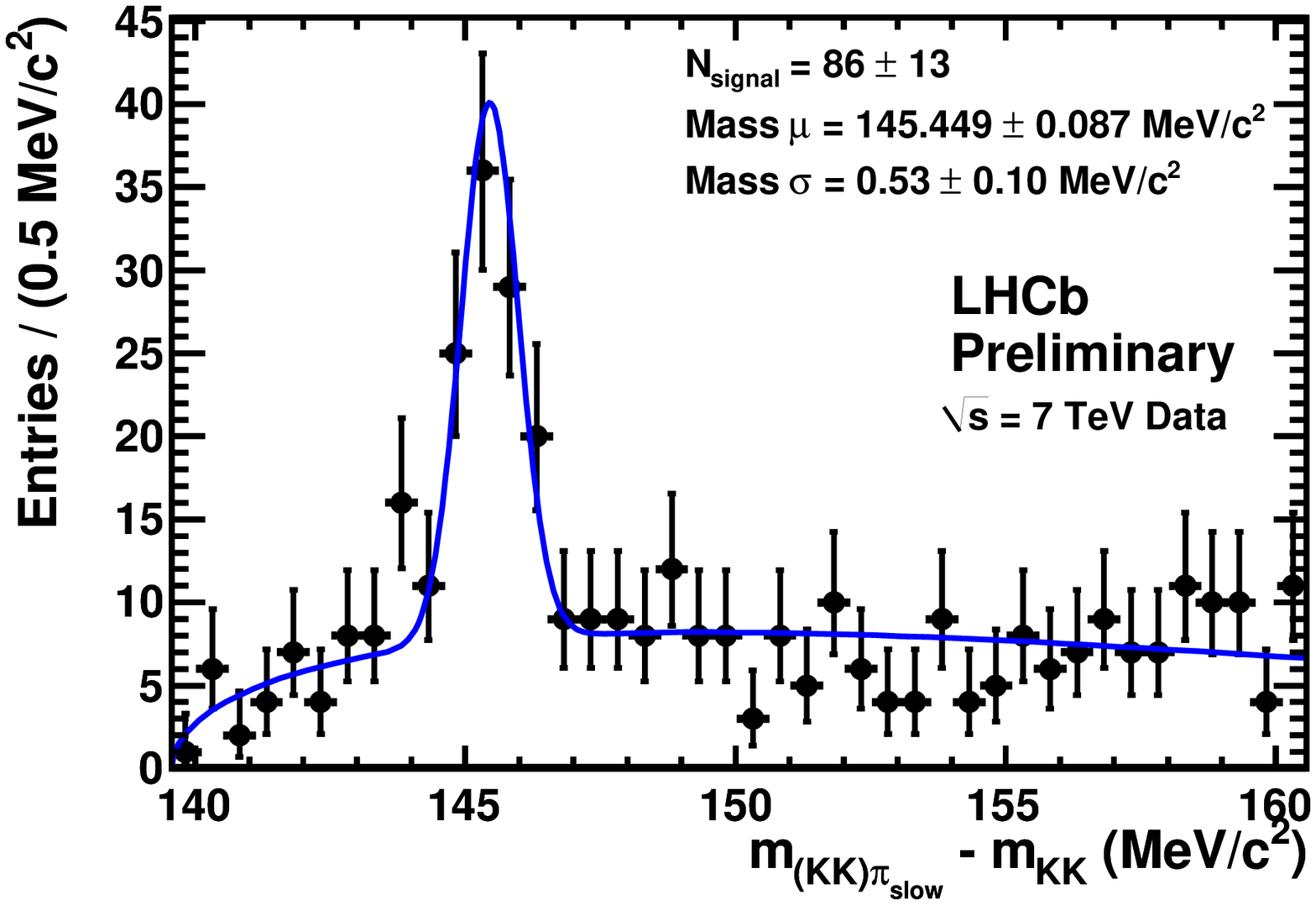,width=0.49\textwidth}
\epsfig{file=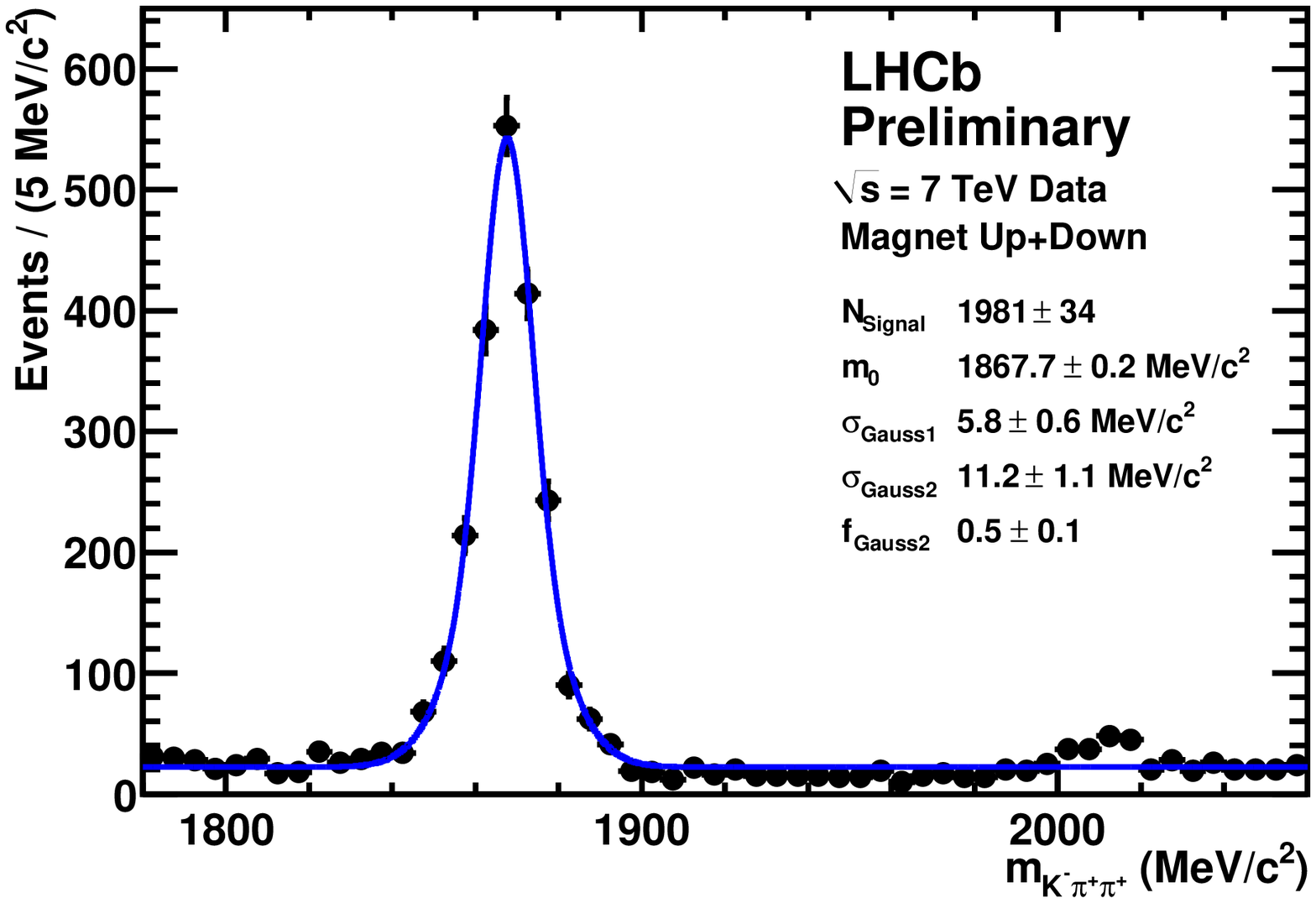,width=0.49\textwidth}
\epsfig{file=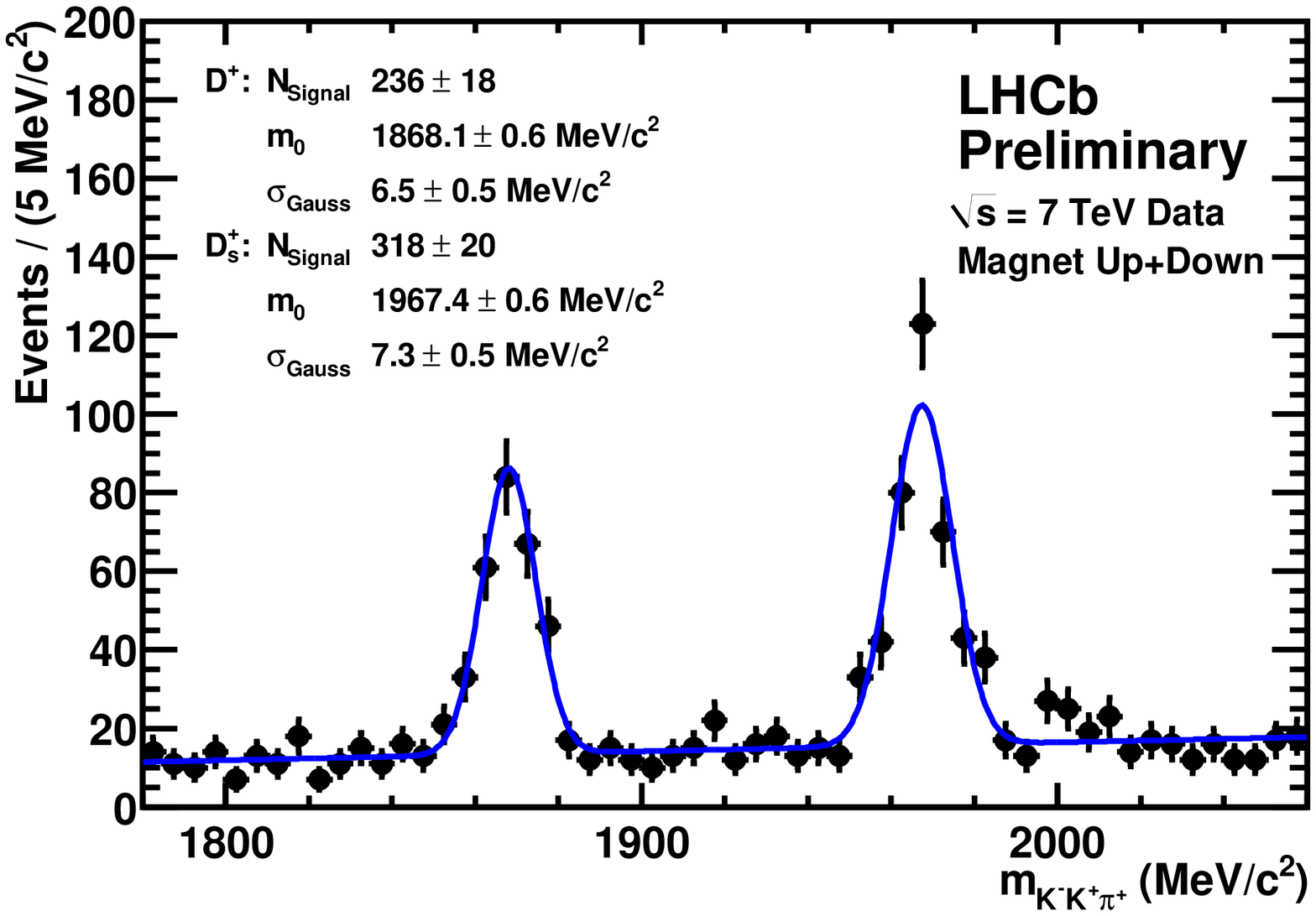,width=0.49\textwidth}
\epsfig{file=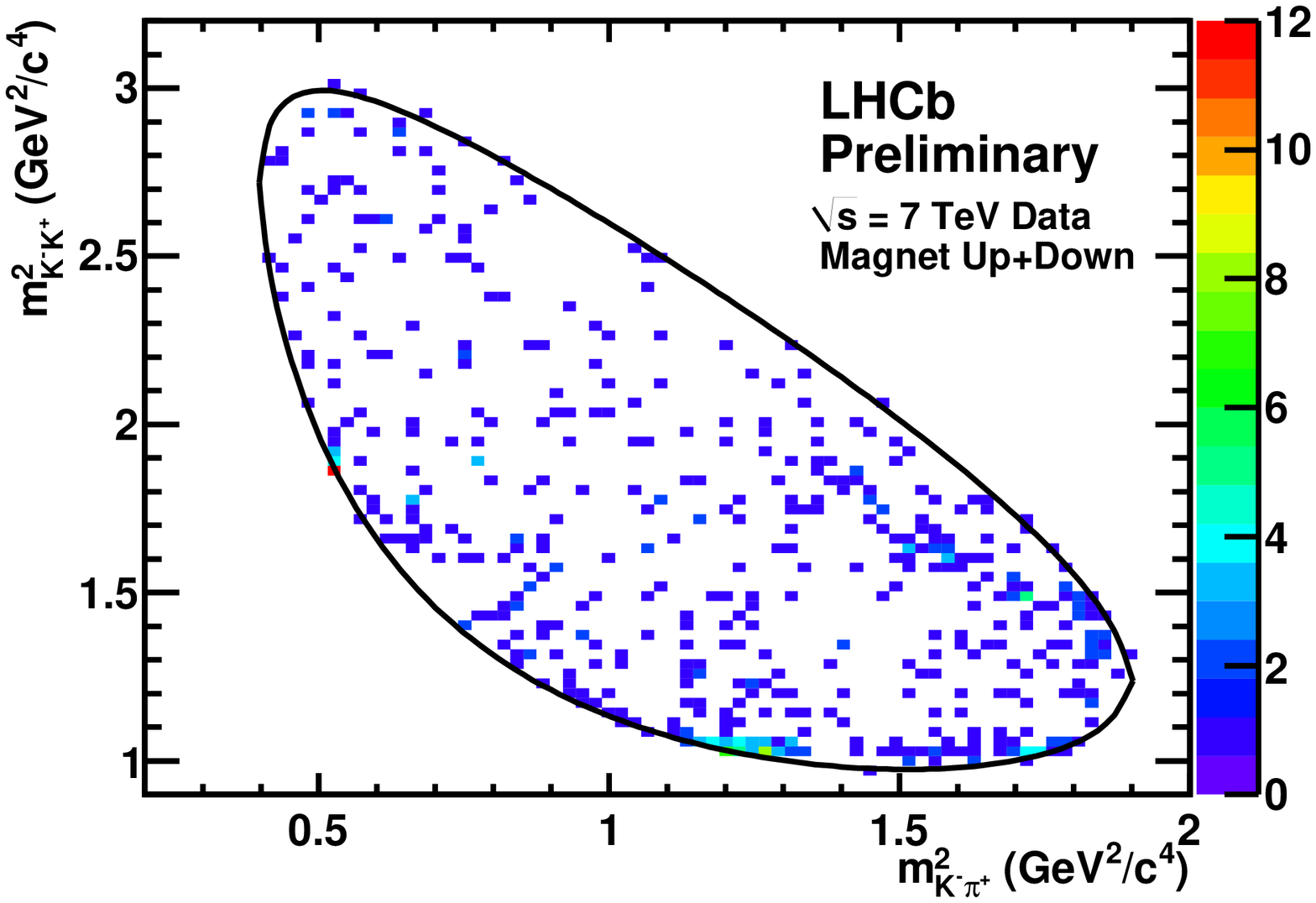,width=0.49\textwidth}
\caption{Some charm signals reconstructed in $\sim 2.7~{\rm nb}^{-1}$
of data at $\sqrt{s}=7$~TeV.
\underline{Top:}
$D^0\to K^-\pi^+$ mass (left) and $D^0\to K^-K^+$ (right) mass.
\underline{Middle:}
difference between the $K^-K^+\pi^+$ and $K^-K^+$ masses for 
$D^{*+}\to D^0\pi^+ \to K^-K^+\pi^+$ candidates (left), 
and $D^+\to K^-\pi^+\pi^+$ mass (right).
\underline{Bottom:}
$K^-K^+\pi^+$ mass showing the $D^+\to K^-K^+\pi^+$ and 
$D_s^+\to K^-K^+\pi^+$ signals (left), and Dalitz plot of
$D^+\to K^-K^+\pi^+$ candidates (right).}
\label{fig:charm}
\end{center}
\end{figure}

Clean charm signals reconstructed in the first $\rm 2.7~nb^{-1}$ 
of data at $\sqrt{s}=7~{\rm TeV}$ (Fig.~\ref{fig:charm})
already allow to firm up 
exciting prospects for measurements of $D^0-\bar{D}^0$ mixing 
and CP violation in the charm sector~\cite{Marks_talk}. 
Indeed, with $\rm 0.1~fb^{-1}$ the statistics of (flavour-tagged)
$D^0$ decays are expected to exceed that of the BABAR experiment
by an order of magnitude.  Significant contributions from LHCb 
are expected soon on several mixing-related observables, in particular: 
\begin{itemize}
\item $y_{\rm CP} = \frac{\tau(D^0\to K^-\pi^+)}{\tau(D^0\to K^-K^+)}-1$
from the proper-time measurements of untagged $D^0\to K^-\pi^+$ and
$D^0\to K^-K^+$ decays (Fig.~\ref{fig:charm} top);
\item $A_\Gamma = \frac{\tau(\bar{D}^0\to K^+K^-)-\tau(D^0\to K^-K^+)}%
{\tau(\bar{D}^0\to K^+K^-)+\tau(D^0\to K^-K^+)}$
from the proper-time measurements of flavour-tagged $D^0\to K^-K^+$ decays,
where the flavour of the $D^0$ meson at production ($D^0$ or $\bar{D}^0$) 
is determined from the sign
of the charged pion in the reconstructed $D^{*-}\to D^0 \pi^+$ decay
(Fig.~\ref{fig:charm} middle left);
\item mixing parameters related to the mass and decay-width differences
in the  $D^0-\bar{D}^0$ system, from the time-dependent analysis of
wrong-sign flavour-tagged $D^0 \to K^+\pi^-$ decays (interference between 
doubly-Cabibbo suppressed decays without mixing and Cabibbo-favoured
decays with mixing).
\end{itemize}

Similarly, huge statistics of charged $D$ mesons will allow
an unprecedented search for direct CP violation in charm.
The most interesting modes are the singly-Cabibbo suppressed decays,
governed by gluonic penguin diagrams where New Physics may enter.
The three-body mode $D^+ \to K^-K^+\pi^+$, together with 
the two Cabibbo-favoured decays $D^+_s \to K^-K^+\pi^+$ and
$D^+ \to K^-\pi^+\pi^+$ to be used as control channels,
offers the interesting possibility of a Dalitz plot analysis
where local CP asymmetries can be probed
(Fig.~\ref{fig:charm} middle right and bottom). 

\section{First \boldmath $b\to J/\psi X$ and $b\to D^0 \mu X$ signals}

\begin{figure}[t]
\begin{center}
\raisebox{31ex}{\epsfig{file=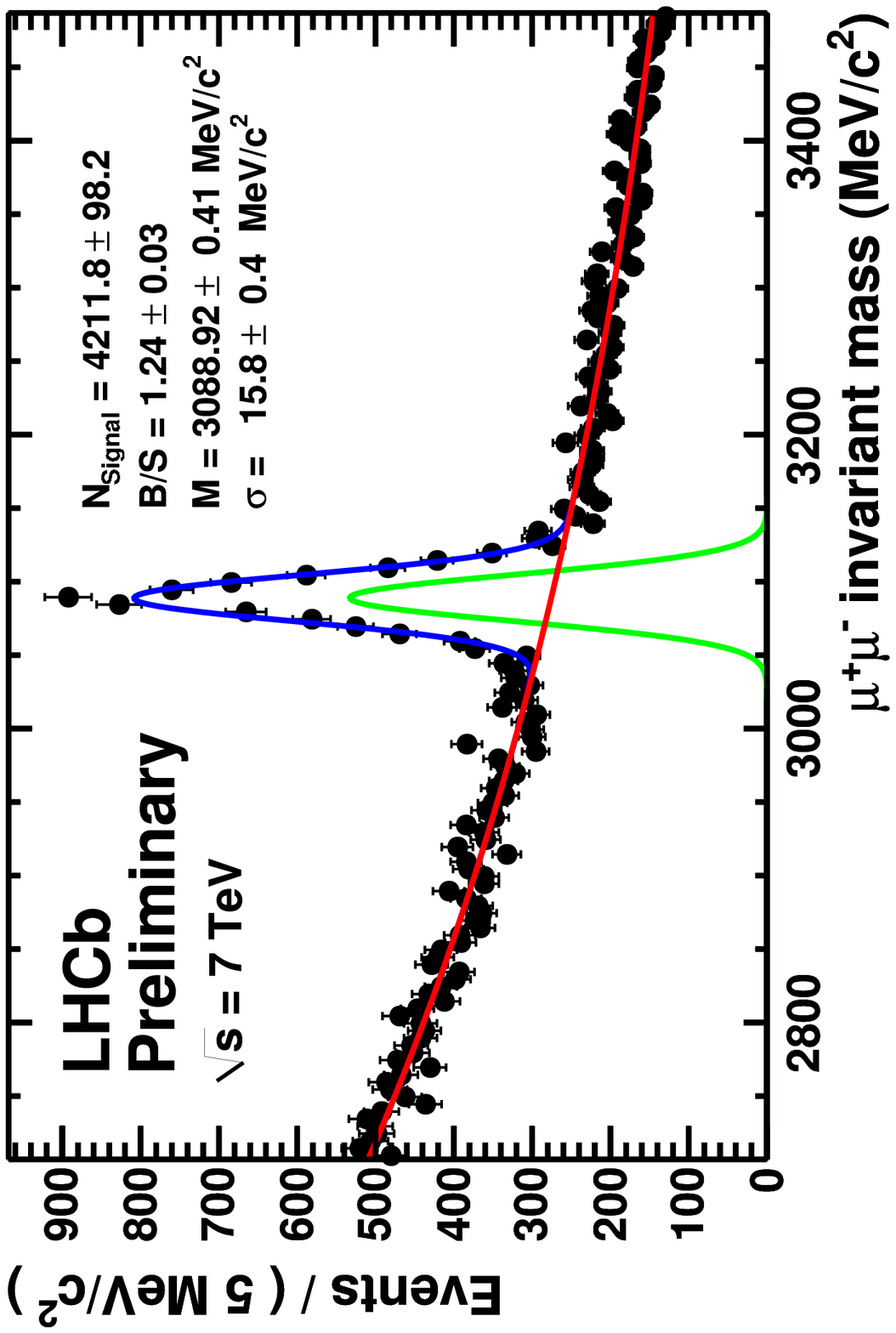,height=0.49\textwidth,angle=270}}
\epsfig{file=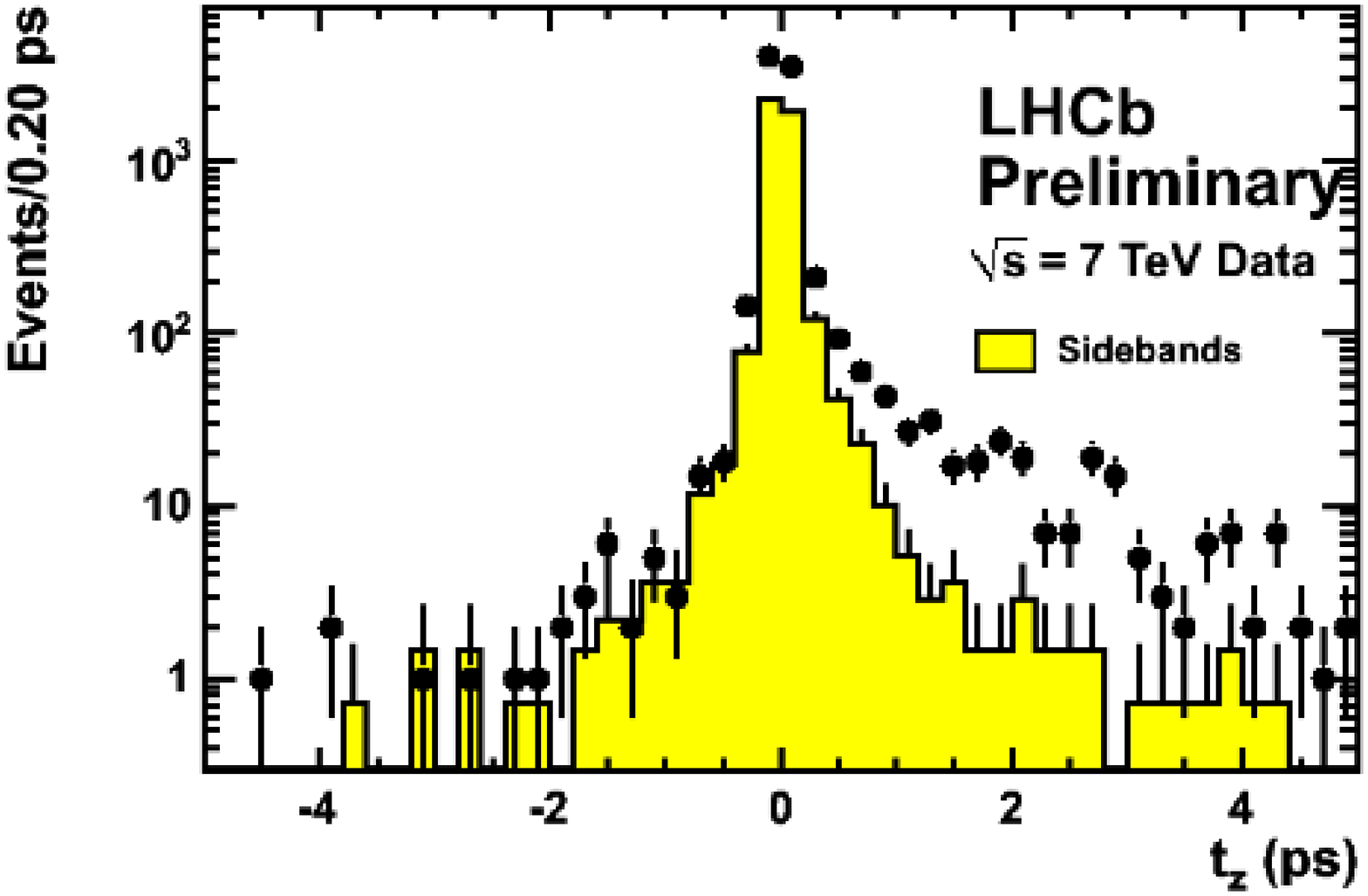,width=0.49\textwidth}
\caption{\underline{Left:} Dimuon invariant mass distribution showing
the $J/\psi\to\mu^+\mu^-$ signal in $\sim 14~{\rm nb}^{-1}$ of data
at $\sqrt{s}=7$~TeV.
\underline{Right:} Pseudo-proper time distribution of the $J/\psi$
candidates in the signal window (black points) and in the sidebands
(yellow histogram). The difference between the two distributions
corresponds to signal $J/\psi$ and displays a tail at large proper time
indicative of $b\to J/\psi X$ production.}
\label{fig:Jpsi}
\end{center}
\end{figure}

\begin{figure}[t]
\begin{center}
\epsfig{file=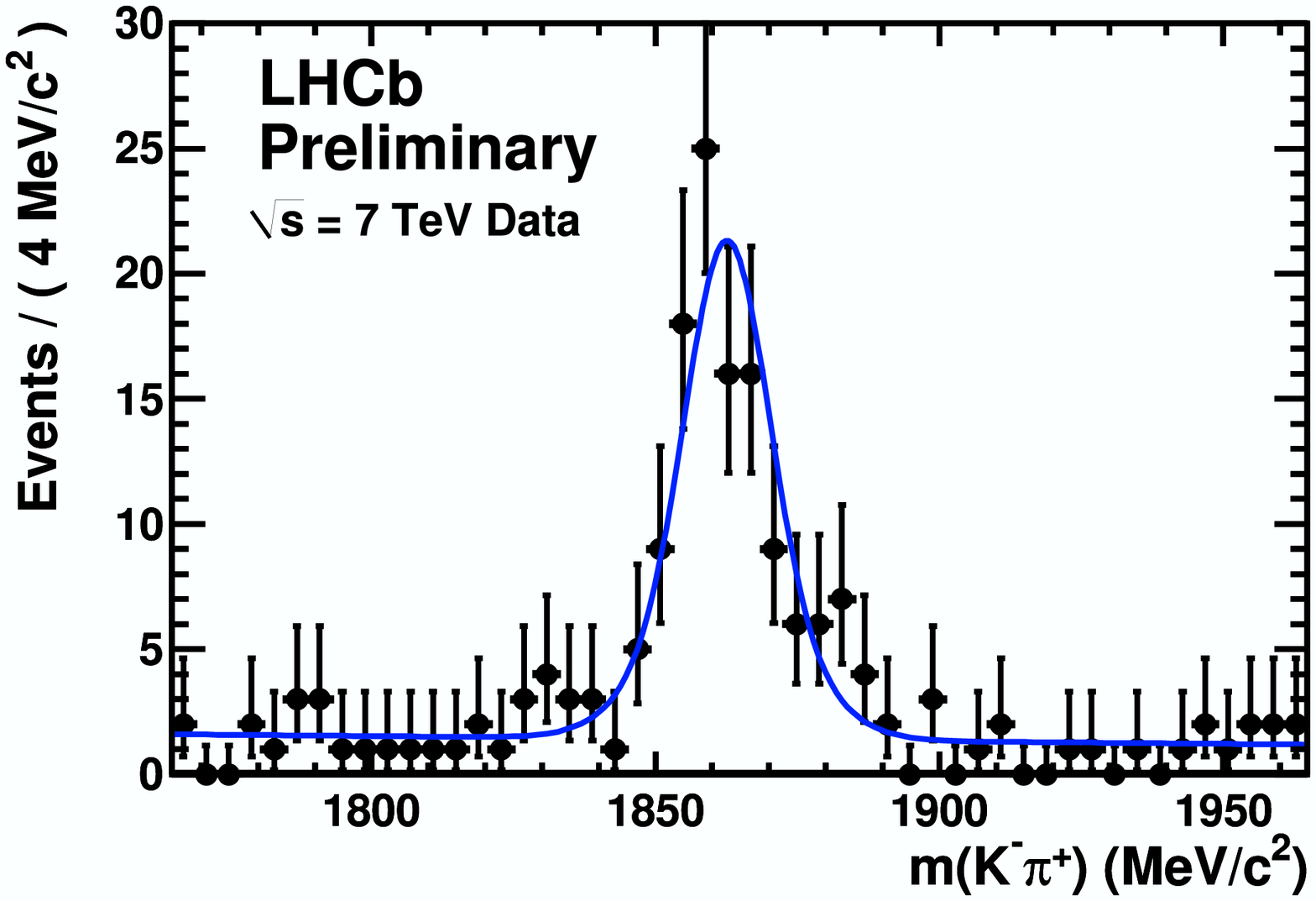,width=0.49\textwidth}
\epsfig{file=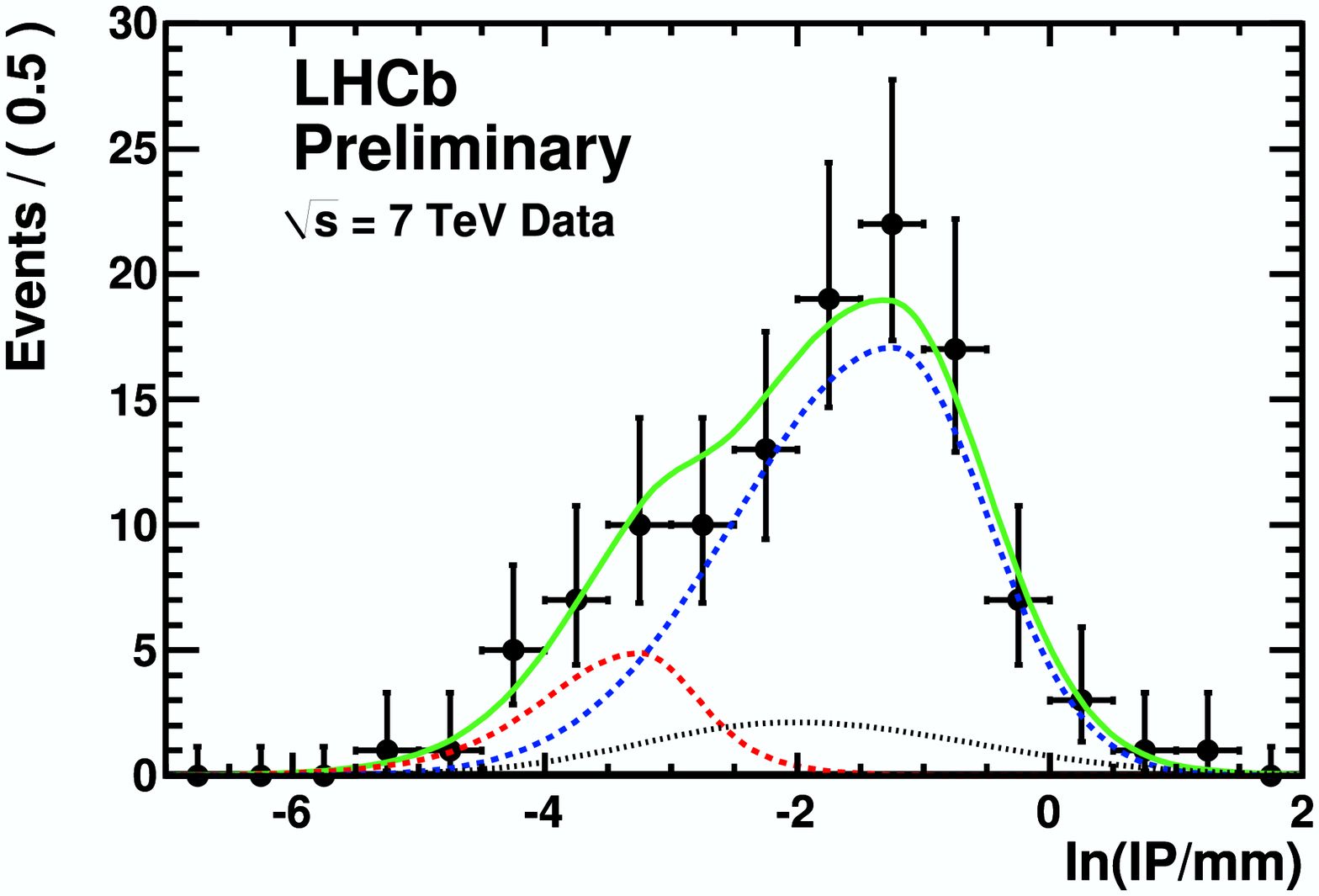,width=0.49\textwidth}
\epsfig{file=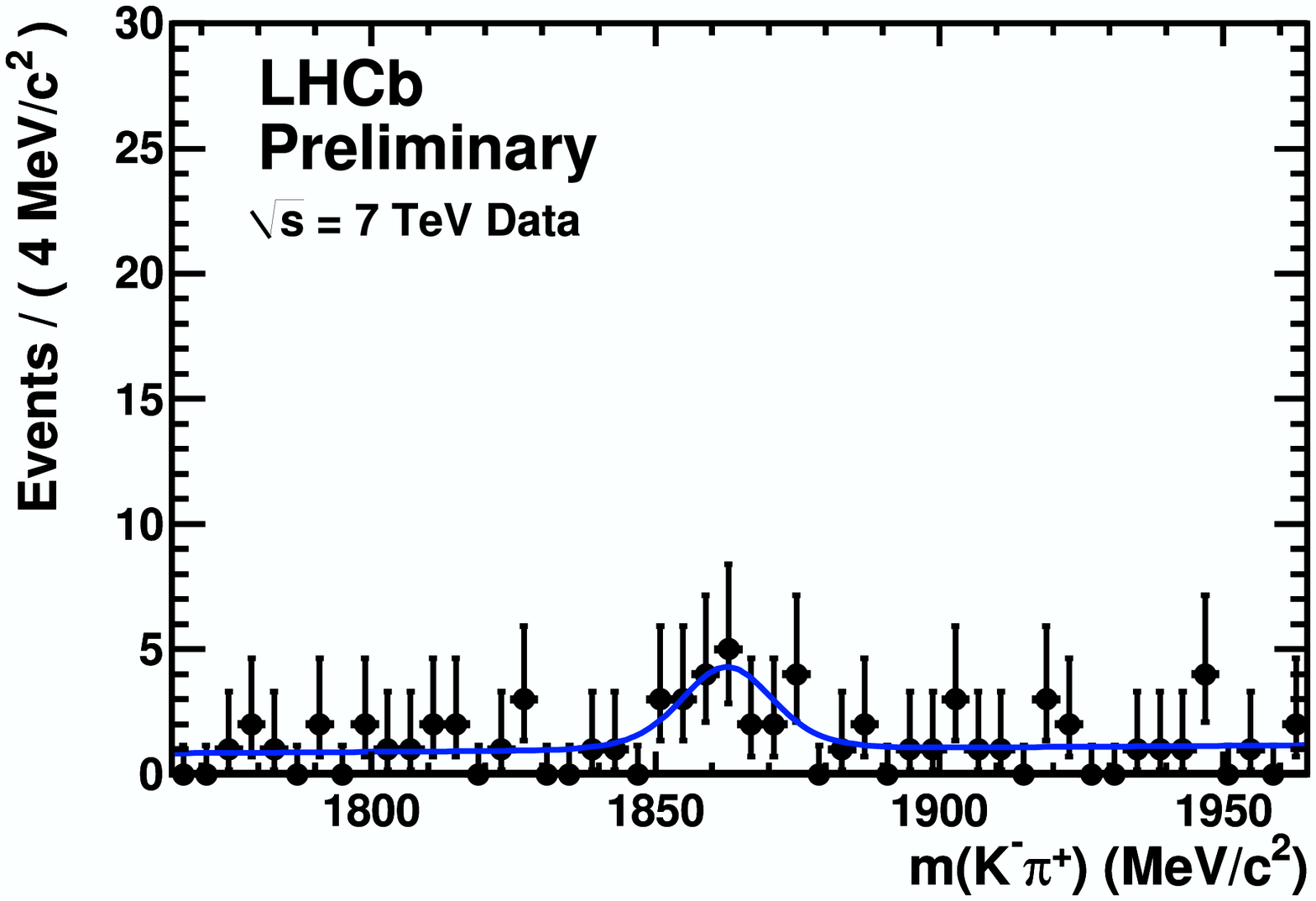,width=0.49\textwidth}
\epsfig{file=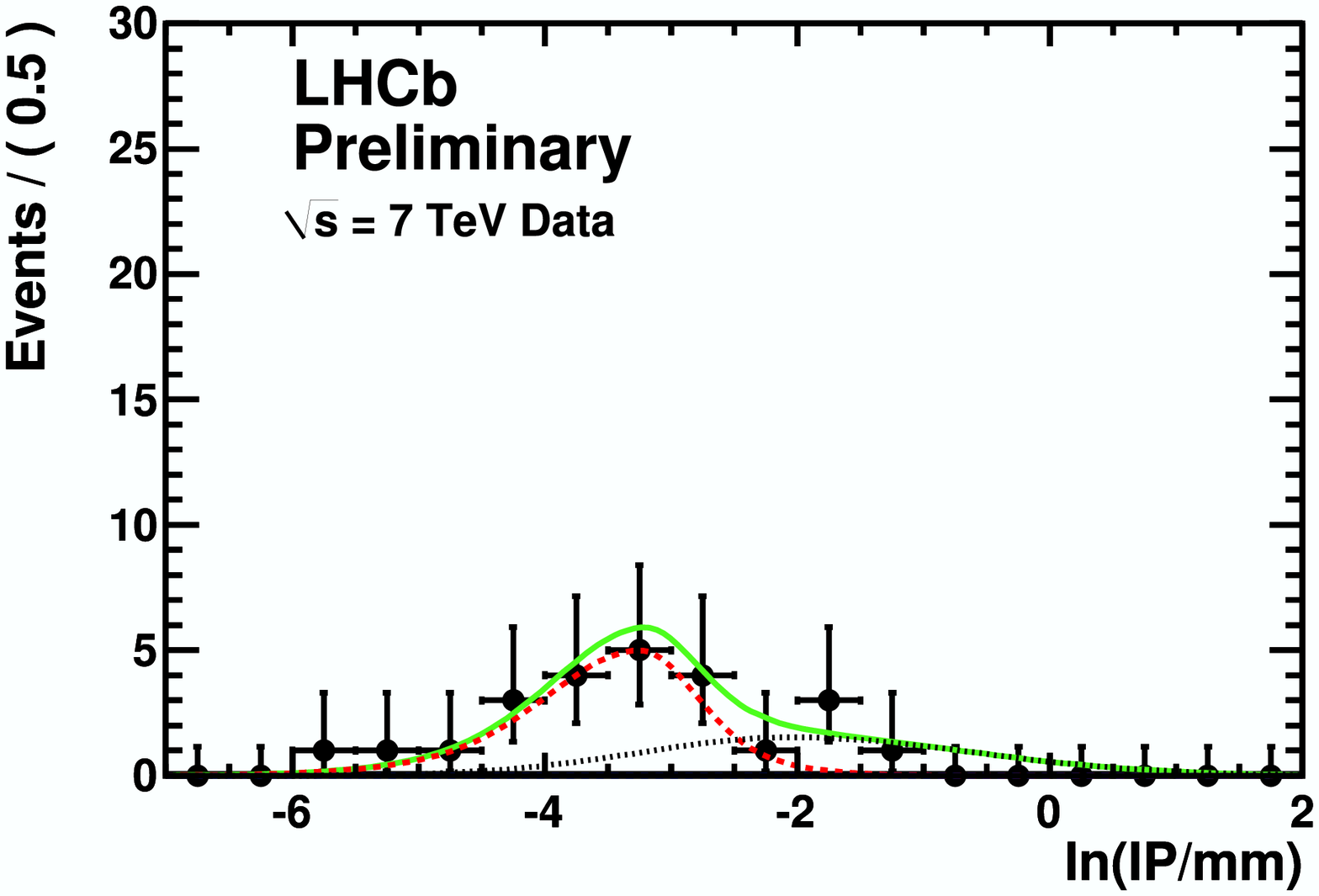,width=0.49\textwidth}
\caption{$D^0\to K^-\pi^+$ invariant mass (left) and logarithm of the $D^0$
impact parameter in millimeters with respect to the primary vertex (right)
for $D^0\mu$ candidates with `right sign' (top) and `wrong sign' (bottom)
correlation, in $\sim 3{\rm nb}^{-1}$ of data at $\sqrt{s}=7$~TeV.
Fit results are superimposed as curves. In the right-hand side plots,
the black dotted curve represents the non-$D^0$ background estimated
from the mass sidebands, and the blue (red) dotted curve represents 
the $D^0$ signal from $b$-hadron decays (prompt production).}
\label{fig:Dzeromu}
\end{center}
\end{figure}

Bottom production can easily be observed with
a few $\rm nb^{-1}$ of data, if inclusive selections are used. 
Two approaches are described here, which will soon yield 
the first measurements of the $b\bar{b}$ production cross section at 
$\sqrt{s}=7~{\rm TeV}$.

An important part of LHCb's physics programme is based on the selection 
of $J/\psi \to \mu^+\mu^-$ decays, which leave a clear signature in the
detector and which can efficiently be recognized both at the trigger level
and in the offline analysis. At the present level of understanding of 
the detector alignment and calibration, and using a very loose trigger, 
a signal of $\sim 300$ events per $\rm nb^{-1}$ is obtained,
with a mass resolution of $16~{\rm MeV}/c^2$ and a signal-to-background 
ratio of 0.8 in a $\pm 45~{\rm MeV}/c^2$ window around the 
central value of the mass peak (Fig.~\ref{fig:Jpsi} left). This abundant 
signal will be an important tool to further understand and 
improve the reconstruction performance. 
The two main sources of $J/\psi$ mesons, prompt production at the $pp$ 
interaction vertex and secondary production in $b$-hadron decays, can be 
separated by measuring the pseudo-proper time
$t_z = (z_{J/\psi}-z_{\rm PV})\times m_{J/\psi}/p_z$,
where $z_{J/\psi}$ and $z_{\rm PV}$ are the reconstructed 
positions of the $J/\psi$ decay and of the $pp$ interaction point along the 
beam direction ($z$ axis), $m_{J/\psi}$ is the nominal $J/\psi$ mass,
and $p_z$ the $z$ component of the reconstructed $J/\psi$ momentum. The 
distribution of $t_z$ is shown in Fig.~\ref{fig:Jpsi} (right) for 
$J/\psi$ candidates with reconstructed masses
in the signal and sideband regions.
The $b \to J/\psi X$ component of the signal is 
clearly visible as an exponential tail in the positive $t_z$ region.

A similar analysis is performed by selecting $D^0 \to K^-\pi^+$ decays and 
using the distribution of the $D^0$ impact parameter (IP) with respect to
the primary vertex to extract the $b$ component. A yield of
$1330 \pm 350\,({\rm stat})$ events is obtained in $\sim 3~{\rm nb}^{-1}$,
which the largest $b$-hadron signal observed so far in LHCb. In order to
increase the purity an identified muon track is required in association
with the $D^0$. If the $D^0\mu$ combination comes from a semileptonic
$b \to D^0 \mu^-\bar{\nu} X$ decay, the muon and the kaon from the
$D^0$ must have equal charges (`right-sign' combination).
Figure~\ref{fig:Dzeromu} shows the distributions of the $D^0$ mass and
of the IP logarithm for both the right-sign and wrong-sign samples.
Prompt $D^0$ production (associated with a random muon) contributes
equally to both samples with small IP values, while semileptonic
$b$-hadron decays contribute with larger IP values only to the
right-sign sample. In the latter a clean and significant ($8\,\sigma$)
signal of $85.3 \pm 10.6\,({\rm stat})$ $b$ events is extracted from
a fit of the $\ln(\rm IP)$ distribution, where the shape of the
$b$ (prompt) component is fixed from MC (data without the muon requirement). 
These results have been finalized~\cite{Dmu_paper} since the conference.

In the future the abundant signals of semileptonic $B^0 \to D^0\mu^+\nu$
and $B^0_s \to D^-_s\mu^+\nu$ decays are expected to play a major role
in the study of CP violation in $B^0$ and $B^0_s$ mixing: Monte Carlo studies
indicate that a measurement competitive with the Tevatron results can be
obtained with less than $1~{\rm fb}^{-1}$ of data, which is the statistics
expected by the end of 2011.

\section{Some prospects with fully reconstructed \boldmath $B$ decays}

\begin{figure}[t]
\begin{center}
\epsfig{file=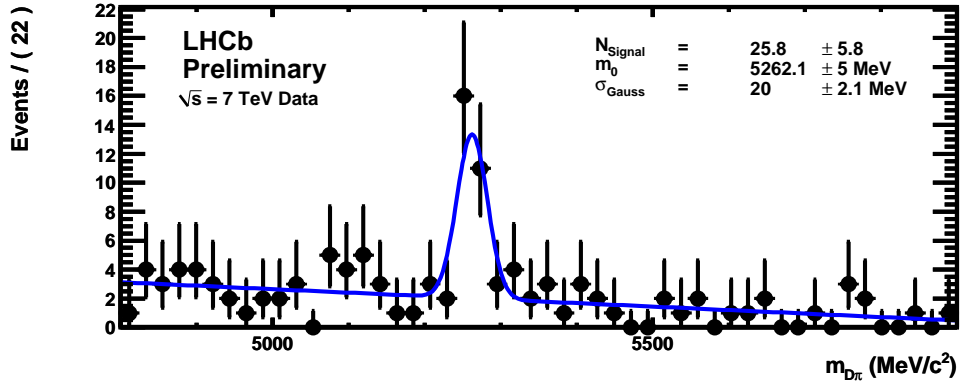,width=0.88\textwidth}
\caption{Sum of the $D^+\pi^-$ and $D^0\pi^+$ invariant mass distributions 
for $\sim 13~{\rm nb}^{-1}$ of data at $\sqrt{s}=7$~TeV, showing
the first signal of exclusively reconstructed $B\to D\pi$ decays.}
\label{fig:Bexcl}
\end{center}
\end{figure}

While several fully reconstructed $B$ candidates have already been selected,
the first significant mass peak has been seen by combining the 
$B^0 \to D^+\pi^-$ and $B^+ \to D^0 \pi^+$ modes (Fig.~\ref{fig:Bexcl}). 
A $B^0_s\to D_s^-\pi^+$ signal as well as $B \to D K$ Cabibbo-suppressed
signals are expected soon. The main physics goal with such hadronic
$B$ decays is the determination of the CKM angle $\gamma$ using the
interference between $b\to c$ and $b\to u$ tree-level diagrams in
$B_{(s)} \to D_{(s)} K$ decays, where a statistical precision of
$\sim 7$~degrees (three times better than the current knowledge
from the $B$ factories) is expected with $1~{\rm fb}^{-1}$
of data~\cite{LHCb:2009ny}.

The current data already allow LHCb to prepare for a few key
$B^0_s$ analyses. Amongst those, the measurement of mixing-induced 
CP violation in $B^0_s \to J/\psi \phi$ decays and the search for the
very rare $B^0_s \to \mu^+\mu^-$ decay based on the first
$0.1~{\rm fb}^{-1}$ of data are expected to compete with Tevatron results,  
and may reveal hints of New Physics with $1~{\rm fb}^{-1}$~\cite{LHCb:2009ny}.

\section{Summary}

LHCb is taking data with success.
First strangeness production measurements have been performed,
and clean charm and bottom signals have been reconstructed. 
LHCb will embark on its core physics programme during the 2010--2011 run,
where the expected integrated luminosity should already give
access to heavy-flavour observables sensitive to possible New Physics.

% ****************************************************************************
% BIBLIOGRAPHY AREA
% ****************************************************************************

% please do not change the following line
\begin{footnotesize}

% please do not change the following line
\end{footnotesize}

% ****************************************************************************
% END OF BIBLIOGRAPHY AREA
% ****************************************************************************

\end{document}